# 2

# U.S.-Singapore cooperation on tech and security: defense, cyber, and biotech*

*Shaun Ee*

*Abstract*

The partnership between the United States and Singapore is founded in no small part on the shared recognition of the value that technology has for national security. Over the last 55 years, Singapore has become an established purchaser of U.S. defense technology, but the past 20 years have also seen the U.S.-Singapore relationship mature into an increasingly collaborative one, tackling newer fields like cybersecurity and biosecurity. However, current geopolitical tensions present a challenge for Singapore, which strives to retain its strategic autonomy by maintaining positive relations with all parties. Paradoxically, the rise of non-traditional security threats may pave the way for greater bilateral cooperation by allowing Singapore to position itself as a hub for cooperation on regional security issues in Southeast Asia at large. In such spirit, this paper recommends that the United States and Singapore do the following: 1) in defense technology, co-develop niche capabilities in C4ISR and unmanned systems with peacetime applications; 2) in cybersecurity, improve their domestic resilience against sophisticated nation-state actors while also building regional capacity to counter cybercrime in Southeast Asia; and 3) in biosecurity, strengthen regional epidemiological surveillance to brace against possible future pandemics.

Shaun Ee

**Introduction**

Since independence in 1965, Singapore has long seen technology as essential for its survival—not just economically, but for defense and security, too. Boxed in geographically, Singapore's government has taken the view that technology is a 'force multiplier' essential to offsetting its spatial constraints.[1] It hence built up its military with acquisitions from other countries—including the United States —in the 1970s and 1980s, even while expanding its defense industry at home. But as Singapore's economy and defense industrial base have matured, it has progressively deepened ties with the United States, growing from an erstwhile client to a steadfast partner. In 2000, the two countries inaugurated the bilateral Defense Cooperation Committee to oversee defense technology cooperation, a testament to that shift in roles.

Twin revolutions have further underscored how science and technology (S&T) cooperation undergirds the two countries' security: first, the information revolution, and second, the pending biotech revolution. In 2015, they signed an Enhanced Defense Cooperation Agreement that identified cyber defense and biosecurity as areas for further cooperation. Washington and Singapore share concerns over the activity of advanced, nation-state-backed threat actors in cyberspace, particularly given an unnamed state's involvement in Singapore's largest data breach in 2018.[2] And Covid-19 has driven home the need for biosecurity cooperation, which will grow in importance as nascent fields like synthetic biology develop.

Bolstering the resilience of their respective societies against malicious actors remains a priority for both the United States and Singapore, such as in cyberspace against advanced nation-state espionage campaigns. But strengthening U.S.-Singapore ties on S&T for security issues does more than improving their respective domestic security—it also provides an opportunity to improve security in the greater Southeast Asian region. As a technologically and economically advanced country known diplomatically as a trusted broker, Singapore is ideal for mediating S&T knowledge flows between the United States and Southeast Asia. Existing regional training programs on cybersecurity are a prime example of this, such as the U.S.-Singapore Third Country Training Program, which helps anchor U.S. presence in the region while building capacity in Southeast Asia.

As the U.S.-Singapore relationship continues to blossom, searching for ways to insulate their own societies from the malicious use of new technologies should remain a focal point of the two countries' collaboration. But further regional capacity building should be an equal priority, as this will help cement the liberal, rules-based international order that both countries are so passionate about defending.

**Defending forward: The roots of bilateral security cooperation**

Singapore's strong defense relationship with the United States is rooted in the strategic realities it faces. As a small country, it depends on technology to 'level the playing field' with its much larger neighbors. But its size also prevents it from achieving defense-industrial sovereignty, forcing it to acquire technology from other major powers for its self-defense. Hence, the United States has played a historically important role in helping Singapore develop its deterrent capacity, although not an exclusive one. In turn, what the United States gained during the Cold War was a partner willing to embrace its continued presence, despite setbacks elsewhere in the region.

Following the Cold War, bilateral cooperation did not halt. Instead, it deepened, with Singapore's continued economic growth and stable political priorities permitting it to invest heavily in next-generation U.S. weapons platforms. Presently, Singapore's considerable force projection capabilities, built in no small part on U.S. equipment, allow it to deter potential aggressors effectively. More than that, however, the growth of its domestic defense industry has enabled it to play a proactive role as a collaborator in the relationship, although the fundamental inequity in size between the United States and Singapore makes true partnership difficult.

*Defense tech cooperation, 1965 – 2000*

The strategic context that Singapore's leaders face now is much the same as the one they faced in 1965: it lacks strategic depth, does not have the population to maintain a sizeable standing army, and depends on much larger neighbors for even basic food and water security. From its

> "…Singapore's considerable force projection capabilities, built in no small part on U.S. equipment, allow it to deter potential aggressors effectively."

leaders' perspective, whether or not Singapore faces extant threats is immaterial because its inherent vulnerabilities expose it to economic and military coercion, if not outright invasion. Hence, from the start of independence, it has maintained the same core philosophy: it must 'invest in defense to deter threats from arising in the first place.'[3]

This survival anxiety has been baked into the thinking of Singapore's leaders since its early days, with its first Prime Minister (PM) Lee Kuan Yew characterizing it as a 'small fish' that could not take its security for granted.[4] Initially, Singapore had hoped for several years to prepare itself for independence under the aegis of remaining British forces, but these hopes were quashed by the collapse of British power worldwide.[5] Facing an accelerated timetable for British withdrawal, it discreetly sought Israeli military advice, and on that basis, rapidly built up a conscript army through the 1960s and 1970s. This formed the core of its professed 'poisonous shrimp' strategy: in the words of then-PM Lee in 1966, making itself unpalatable enough that it would be 'left alone.'[6]

---

[1] Huxley, Tim. *Defending the Lion City: The Armed Forces of Singapore*, The Armed Forces of Asia (St. Leonards, NSW Australia: Allen & Unwin, 2000).
[2] This was the 2018 SingHealth breach, in which 1.5 million patients' data was stolen.
[3] Manohara, Chinniah "Defense Procurement and Industry Policy—A Singapore Perspective," *Defense and Peace Economics* 9, no. 1–2 (March 1998): 119–36, https://doi.org/10.1080/10430719808404897.

[4] Lee, Kuan Yew. "Big and Small Fishes in Asian Waters" (National Archives of Singapore, June 15, 1966), https://www.nas.gov.sg/archivesonline/data/pdfdoc/lky19660615.pdf.
[5] Huxley, *Defending the Lion City*.
[6] Lee, "Big and Small Fishes in Asian Waters."

Yet, by the 1980s, Singapore's defense strategy was already maturing beyond this basic deterrence-by-punishment approach. Publicly, it still professed adherence to the 'poisonous shrimp' doctrine; more privately, it was acquiring force projection capabilities that permitted it to take an offensive, pre-emptive approach.[7] This approach, later dubbed the 'porcupine' strategy, relied increasingly on technology as a 'force multiplier' to offset Singapore's geographical constraints. Through the 1970s and 1980s, Singapore acquired air defense, early warning, and long-range strike capabilities that made up for its nonexistent hinterland. In the late 1980s, it turned to C3I capabilities (i.e., command, control, communications, and intelligence) to give itself a 'strategic edge.'[8]

Particularly in the air domain, the United States became Singapore's chosen supplier for many new capabilities. This choice was likely owed in part to the high quality and logistical support for U.S. equipment, but geopolitics may have been an added consideration. In the shadow of U.S. losses in Southeast Asia in the 1960s and 1970s, Singapore made a point to expedite U.S. regional presence by, for example, allowing it to use Tengah Airbase in the 1970s to conduct patrol flights in the Indian Ocean.[9]

However, Singapore's preference for continued U.S. engagement did not translate into an unqualified endorsement of U.S. platforms. Despite reliance on the United States for cutting-edge aircraft, Singapore has typically turned to European countries like France, Germany, and Sweden for naval capabilities.[10] It has also sought to improve its strategic autonomy by tasking domestic defense manufacturers with the production of smaller platforms and small arms. During occasional rocky spells in the U.S.-Singapore relationship, this willingness to hedge has paid dividends: for example, in the mid-1980s, when the United States refused to help upgrade Singaporean C-130s for aerial refueling, Israel instead stepped in to do the job.[11]

Overall, however, Singapore remained largely enthusiastic about U.S. equipment and U.S. engagement in the region. In 1990, Singapore signed a Memorandum of Understanding (MOU) with the United States, offering up the use of Singapore bases in 1990 following the departure of U.S. forces from Subic Naval Base and Clark Air Base in the Philippines. Later, it even customized Changi Naval Base on its own dime to be capable of housing an aircraft carrier, despite Singapore not having one of its own.[12] By the late 1990s, Singapore was spending up to US$2.5 billion per year, or two-thirds of its defense budget, on infrastructure and procurement—half of which went to foreign equipment and services. Within this, the largest source of foreign military sales was the United States, paving the way for deeper cooperation in 2000 and beyond.[13]

*Defense tech cooperation, 2000 – present*

The turn of the millennium marked a new phase for both U.S.-Singapore defense cooperation and the Singapore Armed Forces at large. Central to this was the maturation of Singapore's domestic defense industry, because it allowed Singapore to assume a role of not just purchaser, but also of partner with the United States.

From a domestic standpoint, the most significant change in defense procurement was the March 2000 establishment of the Defense Science and Technology Agency (DSTA) as the 'executive agent' of Singapore's Ministry of Defense (MINDEF), handling defense acquisition and managing defense research and development.[14] DSTA was carved out from MINDEF's Defense Technology Group (DTG), replacing a structure that was, in the words of DSTA's first director of procurement, 'not nimble and responsive enough to meet the challenges of the future.'[15] As a distinct legal entity, DSTA has greater autonomy than its predecessor to implement defense policy on MINDEF's behalf, which has helped Singapore meet the modernization requirements of the third-generation (3G) SAF.

In parallel to this repositioning of defense policy, Singapore also deepened defense tech cooperation with the United States by creating new forums and institutions. In 2000, it inaugurated the Defense Cooperation Committee (DCC), an annual bilateral forum at the Permanent Secretary/Undersecretary level to oversee defense tech cooperation between the two countries.[16] The DCC oversees 8 out of the 9 bilateral dialogues on U.S.-Singapore technology cooperation that are active as of 2019, although information is limited about the content and nature of these dialogues.[17] Likewise, information on U.S.-Singapore cooperation at the agency-to-agency level is generally not publicly available.

The 2000s also saw the United States increasingly recognize Singapore as a hub for research with national security applications. Alongside the DCC's creation, the

---

Office of Naval Research Global (ONR Global) opened a branch in Singapore in 2000, later staffed up to a full office in 2006.[18] ONR Global monitors and funds research in emerging technologies of interest to the U.S. Navy, although its scope of work in Singapore remains relatively modest. In 2013, it had four in-country experts and supported about $500,000 in Singapore-based projects, with the majority of that going directly toward research grants.[19] ONR Global Singapore also works with ONR Global Tokyo to fund projects in other Asian countries. In 2013, projects totaled $4.02 million across the Area of Responsibility of what was then the U.S. Pacific Command (now the U.S. Indo-Pacific Command).

However, U.S. defense tech sales to Singapore have continued to dwarf these efforts. Between 2014 and 2019 alone, the United States authorized US$37.6 billion in defense sales to Singapore under the Direct Commercial Sales scheme, with an additional US$7.34 billion government-to-government sales cases being active under the Foreign Military Sales scheme.[20] For context, the country's overall defense budget allocation in 2020 was $10.8 billion.[21] As true of the pre-2000s period, aircraft-related sales were a large slice of this, with the two largest deals of the 2010s being a $2.43 billion upgrade of Singapore's F-16s in 2014, and a $2.75 billion order of F-35s in 2020.[22]

Major aircraft deals like these will undoubtedly be a continued feature of Singapore's partnership with the United States, but other trends—such as Singapore's falling birth rate, which promises to substantially reduce its manpower pool for conscription—mean that Singapore is also being forced to think more creatively about how to equip its forces. In the face of what could be a one-third decline in eligible conscripts from the late 2010s to 2030, the country is increasingly exploring unmanned and other capabilities to augment its firepower.[23] The development of these advanced technological capabilities could provide a focal point for U.S.-Singapore collaboration in the future.

**The changing regional security environment, 2000 – present**

Beyond Singapore coming into its own, another decisive factor shaping its relationship with the United States has been the evolving global strategic environment. The end of the Cold War set the stage for a more complex threat environment, featuring the coexistence of hybrid warfare alongside great-power conflict, and state actors alongside non-state actors. U.S. strategy in Southeast Asia has undergone two major shifts in the 21st century—first, the War on Terror and second, the Pivot to Asia—but while the former shift provided Singaporean policymakers with straightforward opportunities for closer cooperation, the latter shift has been more complex for Singaporean policymakers to navigate.

The prospect of worsening U.S.-China relations has led Singapore's leaders to express concern over Asian countries being forced to choose between the two major powers, with among the most public expressions of concern being an article published by PM Lee Hsien Loong in the leading magazine *Foreign Affairs*.[24] While Singapore has remained receptive to strengthening U.S.-Singapore ties, such concerns mean that it is also cautious about taking actions that could prove regionally destabilizing. Nonetheless, the peacetime applications of certain defense technologies—such as artificial intelligence (AI) for disaster relief—present an opportunity for the two countries to collaborate in a measured way.

> **"While Singapore has remained receptive to strengthening U.S.-Singapore ties, such concerns mean that it is also cautious about taking actions that could prove regionally**

*Post-2000s strategic priorities: the war on terror and the pivot to asia*

Through the first decade of the 21st century, the specter of 9/11 haunted both countries and spurred them to work together beyond conventional security ties. In 2003, they affirmed a shared interest in counterterrorism and nonproliferation; then, in 2005, they followed on by signing a Strategic Framework Agreement (SFA) that brought bilateral work on both traditional and non-traditional security work under the same umbrella.[25] Beyond expanding the scope of their cooperation, the 2005 SFA also elevated their relationship by recognizing Singapore as a 'Major Security Cooperation Partner' of the United States, a unique designation that, according to PM Lee Hsien Loong, only Singapore had received as of 2018.[26]

Following the high-level 2005 SFA, the two countries rolled out a number of other homeland security initiatives. In 2007, their respective homeland security agencies signed an agreement to collaborate on technology for counterterrorism, cybersecurity, and other purposes.[27] In 2009, the United States established a regional office of its

---

[18] Vu, Cung. "Office of Naval Research Global," https://sites.nationalacademies.org/cs/groups/pgasite/documents/webpage/pga_147331.pdf.
[19] Ibid. This total is across three ONR Global programs: the Collaborative Science Program, which supports seminars and workshops; the Visiting Scientist Program, which funds travel of non-U.S. scientists to the United States; and the Naval International Cooperative Opportunities Program, which provides direct funding for research projects by international scientists. Expenditure across these three programs totaled about $13.7 million in 2013.
[20] Bureau of Political-Military Affairs, "U.S. Security Cooperation With Singapore."
[21] Parameswaran, Prashanth. "What Does Singapore's New Defense Budget Say About the Country's Security Thinking?," *The Diplomat*, February 24, 2020, https://thediplomat.com/2020/02/what-does-singapores-new-defense-budget-say-about-the-countrys-security-thinking/.
[22] U.S. Defense Security Cooperation Agency. "Singapore - F-16 Block 52 Upgrade," January 14, 2014, https://www.dsca.mil/press-media/major-arms-sales/singapore-f-16-block-52-upgrade. Reuters Staff, "U.S. State Dept. Approves Sale of 12 F-35 Jets to Singapore," *Reuters*, January 10, 2020, https://www.reuters.com/article/us-singapore-defence-lockheed-idUSKBN1Z90G9.
[23] Ungku, Fathin, and Miyoung Kim, "Singapore Armed Forces Going More Hi-Tech as Recruiting Levels Seen Sliding," *Reuters*, June 30, 2017, https://www.reuters.com/article/us-singapore-defence-idUSKBN19L19R.
[24] Lee, Hsien Loong "The Endangered Asian Century," *Foreign Affairs*, December 18, 2020, https://www.foreignaffairs.com/articles/asia/2020-06-04/lee-hsien-loong-endangered-asian-century.
[25] Singapore Ministry of Defense. "Factsheet - The Strategic Framework Agreement," July 12, 2005, National Archives of Singapore, https://www.nas.gov.sg/archivesonline/data/pdfdoc/MINDEF_20050712001/MINDEF_20050712003.pdf.
[26] Lee, Hsien Loong. "Remarks by PM Lee Hsien Loong at the Joint Press Engagement with U.S. VP Mike Pence" (Prime Minister's Office, Singapore, January 18, 2019), https://www.pmo.gov.sg/Newsroom/remarks-pm-lee-hsien-loong-joint-press-engagement-us-vp-mike-pence.
[27] U.S. Department of Homeland Security and Singapore Ministry of Home Affairs. "Agreement Between the Government of the United States of America and the Government of Singapore on Cooperation in Science and Technology for Homeland/Domestic Security Matters," March 27, 2007, https://www.dhs.gov/xlibrary/assets/agreement_us_singapore_sciencetech_cooperation_2007-03-27.pdf.

Defense Threat Reduction Agency (DTRA), supporting the efforts of the United States throughout Asia to reduce the risk from chemical, biological, radiological, and nuclear weapons.[28] However, even as these initiatives were being implemented, a different set of geopolitical forces were entering into play, requiring the recalibration of the U.S.-Singapore relationship.

The early 2010s saw tensions begin to ratchet up in U.S.-China relations, with China's expanded international footprint under Hu Jintao met with the Obama administration's corresponding Pivot to Asia. Since then, U.S.-China relations have further deteriorated, placing Singapore in a difficult position given its stated preference for an inclusive regional architecture and its strong trade and investment ties with both countries. In a *Foreign Affairs* article, PM Lee Hsien Loong has publicly cautioned that should U.S.-China frictions worsen to the point that Singapore and other Asian countries are forced to pick sides, the region's future prospects could be placed in jeopardy.[29]

Nonetheless, PM Lee has expressed that Asian countries regard the United States as a 'resident power' in Asia, and Singapore has been receptive toward overtures to strengthen U.S.-Singapore relations through the 2010s. In 2012, with the launch of a new annual dialogue, the U.S.-Singapore relationship 'moved… up a weight class' to become a strategic partnership.[30] The same year, the two countries agreed that the United States could deploy up to four Littoral Combat Ships (LCSs) to Singapore on a rotational basis and without basing arrangements.[31] The LCS program's rollout has been troubled, with delays resulting in only three single-ship deployments by 2018, due in part to an unreliable propulsion system and weathering from the long journey between continental United States and Singapore, but U.S. Navy officials have expressed a continued intent to consolidate the program.[32]

In 2015, which marked the 25th anniversary of the 1990 MOU, the two countries further deepened security ties by signing an Enhanced Defense Cooperation Agreement (Enhanced DCA). Combined with other changes, this led one Singaporean official to assess in 2015 that the U.S.-Singapore defense relationship looked 'qualitatively different than… just five years ago.'[33] The 2015 Enhanced DCA built on the previous DCA included in the 2005 SFA, laying out a framework that included technology as one of five key areas for cooperation.[34] Beyond these core five areas, it also highlighted other new arenas for cooperation: humanitarian assistance and disaster relief (HA/DR), cyber defense, biosecurity, and public communications.

*Future outlook and fundamental constraints*

The Enhanced DCA's expanded ambit is particularly important because it provides additional avenues for U.S.-Singapore security cooperation beyond conventional defense. While Singapore has publicly affirmed the value of a continued U.S. presence in the Pacific, it has also been emphatic about the limits of its cooperation with the United States as a partner rather than a formal ally.[35] Singapore is unlikely to engage in joint defense endeavors that might worsen the regional security environment or compromise its claim to neutrality.[36] For example, when the U.S. Secretary of the Navy lofted the idea of stationing a 'First Fleet' out of Singapore in late 2020, Singapore was quick to direct attention to the 2012 agreement on LCS deployments as the 'standing arrangement' between the two countries.[37]

> "Developing these niche high-tech tools allows the United States and Singapore to keep their collaboration subdued but still substantive."

Looking to the future, this means that dual-use technology with peacetime applications could be one of the most promising avenues for cooperation, as seen from a 2019 bilateral agreement on AI applications for HA/DR.[38] Within HA/DR, numerous AI applications exist, ranging from computer vision in remote sensing to robotic autonomy in hazardous terrain.[39] These are all valuable in peacetime, particularly given neighboring Indonesia's susceptibility to natural disasters, and have potential value to both countries' militaries. Developing these niche high-tech tools allows the United States and Singapore to keep their collaboration subdued but still substantive.

**The homeland angle: cybersecurity and biosecurity**

Aside from the above security concerns, however, technological advances since the 2000s have also introduced a new series of threats. Here, ironically, the growing complexity of the threat environment may make it easier for Singapore and the United States to find ways to collaborate. While there is a limit on how closely the United States and Singapore can collaborate on conventional defense technology, security solutions in cybersecurity and biosecurity are inherently multilateral. Rather than doubling down in defense technology, a particularly promising path forward for U.S.-Singapore S&T security cooperation appears to be branching out into other areas.

---

[28] "Defense Threat Reduction Agency (DTRA)" U.S. Embassy in Singapore, accessed April 18, 2021, http://sg.usembassy.gov/embassy/singapore/sections-offices/defense-threat-reduction-agency-dtra/.
[29] Lee, "The Endangered Asian Century."
[30] Adelman, David. "The U.S.-Singapore Strategic Partnership: Bilateral Relations Move Up a Weight Class," *The Ambassador's Review*, 2012. At the time of this comment, Adelman was the serving U.S. Ambassador to Singapore.
[31] Kuok, "The U.S.-Singapore Partnership." Ristian Atriandi Supriyanto, "U.S. Pivots to Maritime Southeast Asia," in *The South China Sea Disputes*, by Yang Razali Kassim (World Scientific, 2017), 109–12, https://doi.org/10.1142/9789814704984_0025.
[32] Larter, David. "U.S. Navy Prepares Major Surge of Littoral Combat Ship Deployments," *Defense News*, July 31, 2020, sec. Naval, https://www.defensenews.com/naval/2020/07/31/the-us-navy-is-preparing-a-major-surge-of-lcs-deployments/. Dzirhan Mahadzir, "CNO: U.S. Still Committed to Littoral Combat Ship Deployments in Southeast Asia," *USNI News*, November 1, 2018, sec. News & Analysis, https://news.usni.org/2018/11/01/cno-u-s-still-committed-littoral-combat-ship-deployments-southeast-asia.
[33] Parameswaran, Prashanth. "Strengthening U.S.-Singapore Strategic Partnership: Opportunities and Challenges," *RSIS Commentary* (blog), August 8, 2016, https://www.rsis.edu.sg/rsis-publication/rsis/co16201-strengthening-us-singapore-strategic-partnership-opportunities-and-challenges/.
[34] Kuok, "The U.S.-Singapore Partnership." Singapore Ministry of Defense, "Singapore, U.S. Step Up Defence Cooperation," December 8, 2015, /web/portal/mindef/news-and-events/latest-releases/article-detail/2015/december/2015Dec08-News-Releases-02572. The five key areas outlined in the framework were military, policy, strategy, technology, and non-conventional security matters (including piracy and terrorism).
[35] Kuok, "The U.S.-Singapore Partnership."
[36] Lee, "The Endangered Asian Century." Cortez Cooper and Michael Chase, *Regional Responses to U.S.-China Competition in the Indo-Pacific: Singapore* (RAND Corporation, 2020), https://doi.org/10.7249/RR4412.5.
[37] Singapore Ministry of Defense. "Reply to Queries on U.S. SECNAV's Calls for New U.S. 1st Fleet Out of Singapore," November 18, 2020, https://www.mindef.gov.sg/web/portal/mindef/news-and-events/latest-releases/article-detail/2020/November/18nov20_mq.
[38] U.S. Department of Defense. "JAIC and DSTA Forge Technology Collaboration," June 27, 2019, https://www.defense.gov/Newsroom/Releases/Release/Article/1888859/jaic-and-dsta-forge-technology-collaboration/.
[39] Gupta, Ritwik. "SEI Podcast Series: AI in Humanitarian Assistance and Disaster Response," interview by Andrew Mellinger, September 2019, https://resources.sei.cmu.edu/library/asset-view.cfm?assetid=634757.

*Cybersecurity: the state of play*

Both countries have good reason to embrace international cybersecurity collaboration, as they have been directly impacted by a slew of cyber incidents in the past several years. By its size and prominence as a target, the United States has suffered more major incidents, ranging from election interference in 2016 to the Baltimore ransomware attack of 2019 to the Sunburst (or SolarWinds) supply chain compromise targeting U.S. government agencies, discovered in early 2021.[40] But neither has Singapore gotten off scot-free: most prominently in 2018, healthcare institutions under the SingHealth umbrella had their systems breached by what was likely a nation-state actor, threatening Singapore's national security by exfiltrating data that included the personal medical information of PM Lee Hsien Loong.[41]

As seen in the Sunburst and SingHealth incidents, sophisticated nation-state espionage campaigns suggest that one priority area for both Singapore and the United States should be building out advanced defensive measures to guard against nation-state espionage. Given the two countries' high level of digitization, cyber-enabled espionage is a particularly attractive vector for other nation-states seeking access to sensitive information about the economic and national security of the United States and Singapore. At the same time, the two countries cannot neglect other varied threat actors such as financially motivated criminal enterprises, or other cyber incidents such as destructive ransomware attacks. Large-scale ransomware attacks, for example, can disrupt vital services and debilitate companies, with the NotPetya attack in 2017 freezing port operations and costing shipping giant Maersk up to US$300 million.[42] Given the cross-border repercussions of such incidents, it is in the interests of the United States and Singapore to bolster the resilience of not only their own societies but also the region's.

*Cybersecurity: existing U.S.-Singapore cooperation*

While formal U.S.-Singapore cooperation on cybersecurity has existed for some time, their joint efforts have accelerated since the mid-2010s, as their respective governments have restructured their institutions to place cybersecurity work under prominent national-level agencies. In 2015, Singapore founded the Cyber Security Agency (CSA), absorbing a range of other agencies from its Ministry of Home Affairs, Infocomm Development Authority, and so on.[43] Not long after, in 2018, the United States elevated its National Protection and Programs Directorate, housed under the U.S. Department of Homeland Security, to the Cybersecurity and Infrastructure Security Agency (CISA), reflecting the independent realization in both countries of the need to empower national agencies to combat these issues.[44]

The two countries signed an MOU on 'Cooperation in the Area of Cybersecurity' in August 2016, the first such agreement between the United States and an Association of Southeast Asian Nations (ASEAN) member-state.[45] Per CSA's press release, the agreement covers a wide range of activities such as incident response coordination, joint cybersecurity exercises, collaboration on regional capacity building, and sharing best practices and information between their respective national Computer Emergency Response Teams (CERTs).[46] This MOU was signed on the sidelines of a larger bilateral meeting between then-U.S. President Barack Obama and Singapore PM Lee Hsien Loong, and so was accompanied by a Singapore-U.S. joint statement committing the two countries to a multi-stakeholder approach to internet governance and a common approach to cyber stability.[47]

Beyond the MOU, another initiative that both countries have pushed is cybersecurity capacity building in Southeast Asia. The region is particularly vulnerable on this front, with the cybersecurity sector lagging the rapidly growing digital economies of countries such as Indonesia and the Philippines. Singapore has turned cybersecurity into a top-tier agenda item in its regional engagements, aggressively pushing both capacity-building and norms-setting during its 2018 chairmanship of ASEAN.[48] In tandem with its diplomatic leadership, it has also proactively invested in regional cybersecurity, such as by launching a S$10-million ASEAN Cyber Capacity Program in 2016 and an ASEAN-Singapore Cybersecurity Center of Excellence in 2019.[49]

For its part, the United States has been keen to support Singapore's efforts. Their cooperation dates back to the 2012 establishment of the Singapore-U.S. Third Country Training Program (TCTP), which aims to improve regional connectivity and resilience by providing workshops on topics ranging from cybersecurity to trade facilitation. By 2017, the TCTP had trained 1,000 participants from ASEAN member states, though these were not exclusively cybersecurity-related trainings.[50] In 2018, the two countries additionally agreed to implement a Singapore-U.S. Cybersecurity Technical Assistance Program for ASEAN member-states, which would deliver three training workshops annually.[51] These joint capacity-building efforts are promising and could provide a model for cooperation in other areas.

---

[40] Though this espionage campaign is most commonly known as the "SolarWinds attack," this article refers to it by the name of one key piece of malware, "Sunburst," since SolarWinds was the company affected and only one of several.

[41] Singapore has not publicly attributed this attack beyond affirming that the responsible party was likely a nation-state. Kevin Kwang, "Singapore Health System Hit by 'Most Serious Breach of Personal Data' in Cyberattack; PM Lee's Data Targeted," *CNA*, July 20, 2018, https://www.channelnewsasia.com/news/singapore/singhealth-health-system-hit-serious-cyberattack-pm-lee-target-10548318.

[42] Leovy, Jill. "Cyberattack Cost Maersk as Much as $300 Million and Disrupted Operations for 2 Weeks," *Los Angeles Times*, August 18, 2017, sec. Business, https://www.latimes.com/business/la-fi-maersk-cyberattack-20170817-story.html.

[43] Tan, Weizhen. "New National Agency to Tackle Cyber Threats," *TODAYonline*, January 28, 2015, https://www.todayonline.com/singapore/new-national-agency-tackle-cyber-threats. Though housed under the Prime Minister's Office, the CSA is managed administratively by the Ministry of Communications and Information. Its current head, David Koh, has a concurrent position at MINDEF. See: Cyber Security Agency of Singapore, *Singapore's Cybersecurity Strategy*, 2016.

[44] Cimpanu, Catalin. "Trump Signs Bill That Creates the Cybersecurity and Infrastructure Security Agency," *ZDNet*, November 16, 2018, https://www.zdnet.com/article/trump-signs-bill-that-creates-the-cybersecurity-and-infrastructure-security-agency/.

[45] Hung, Harry. "Confronting Cybersecurity Challenges through U.S.-Singapore Partnership," *RSIS Commentary* (blog), August 24, 2016, https://www.rsis.edu.sg/rsis-publication/rsis/co16215-confronting-cybersecurity-challenges-through-us-singapore-partnership/.

[46] Cyber Security Agency of Singapore. "Singapore Strengthens Partnership with the United States," August 3, 2016, https://www.csa.gov.sg/news/press-releases/singapore-us-mou.

[47] Office of the Press Secretary. The White House, "Joint Statement by the United States of America and the Republic of Singapore," August 2, 2016, https://obamawhitehouse.archives.gov/the-press-office/2016/08/02/joint-statement-united-states-america-and-republic-singapore.

[48] Parameswaran, Prashanth. "ASEAN Cybersecurity in the Spotlight Under Singapore's Chairmanship," *The Diplomat*, May 2, 2018, https://thediplomat.com/2018/05/asean-cybersecurity-in-the-spotlight-under-singapores-chairmanship/.

[49] Iswaran, S. "Opening Remarks by Mr. S. Iswaran, Minister for Communications and Information, At The ASEAN Ministerial Conference on Cybersecurity" (The Third ASEAN Ministerial Conference on Cybersecurity (AMCC), Singapore, September 19, 2018), https://www.mci.gov.sg/pressroom/news-and-stories/pressroom/2018/9/opening-remarks-by-mr-s-iswaran-at-the-asean-ministerial-conference-on-cybersecurity. Both the $10 million and $30 million sums are intended to be spent over a five-year span.

*Biosecurity: the state of play*

The two countries also identified biosecurity as an area for future cooperation in the 2015 Enhanced DCA, a choice that has proven prescient given the ongoing disruption caused by Covid-19. However, previous unofficial dialogues suggest that experts from both countries differ somewhat in their view of the issue. In a 2014 bilateral dialogue convened by the Johns Hopkins Center for Health Security (CHS), U.S. participants were particularly concerned about deliberate biological attacks, while Singaporean participants indicated that naturally occurring pandemics remained their paramount focus.[52] That perception gap appears to have narrowed in recent years, with Southeast Asian discussants elevating the threat of bioterrorism in the most recent 2019 dialogue, though it remains to be seen how Covid-19 will shape this.[53]

Historically, naturally occurring pandemics have been of greatest concern in Singapore and Southeast Asia—likely the reason that they are top-of-the-mind for regional biosecurity experts. By contrast, it was the 2001 anthrax attacks that highlighted biosecurity threats in the United States.[54] For Asia-Pacific countries, this heightened consciousness has come with certain boons. In all likelihood, one reason that politically diverse countries such as Singapore, Taiwan, and Vietnam could respond effectively to Covid-19 was their prior experience with the severe acute respiratory syndrome (SARS) outbreak from 2002 to 2004.[55] Experiences such as SARS forced many of these countries to put strong pandemic preparedness programs in place and equipped medical specialists with necessary knowledge. However, past experience alone does not guarantee future safety, and the continued role of Southeast Asia as a likely origin point for future pandemics demands vigilance on the part of Singapore and the United States.

Future pandemics are likely to stem from emerging infectious diseases (EIDs)—that is, diseases that are either new or spreading rapidly—and particularly EIDs of zoonotic (i.e., animal) origin, to which humans lack existing immunity.[56] For various reasons, Southeast Asia is a hotspot for such EIDs: it features a diverse range of pathogens, has environmental conditions (e.g., climate) that favor mutation and adaptation of these pathogens, and is home to dense and mobile human populations that regularly interact with animals.[57] As Covid-19 demonstrates, these EIDs can spread quickly across national borders, meaning that even a strong domestic public health infrastructure such as Singapore's does not guarantee protection against EIDs originating in other countries.

Meanwhile, biological incidents with human involvement also remain a real and growing possibility. Such incidents could take several forms, including accidental 'lab leaks' or deliberate bioterror attacks. For now, the chance of such incidents occurring in Southeast Asia remains low because of the limited maturity of regional biotech research, but this may not remain true in the future. Singapore, for example, has announced that it will be building ASEAN's first biosafety level four (BSL-4) lab, which will be equipped to deal with the world's most dangerous pathogens.[58] This will be a boon for research into countermeasures against such pathogens, but as it and other countries continue to construct such facilities, they must ensure that safety measures are followed appropriately.[59]

Particularly concerning for the United States and Singapore is the development of new biotech capabilities that could make it easier for non-state actors to design or manufacture biological weapons. Currently, there are high barriers to doing so, as this depends on unreliable and idiosyncratic techniques that it takes years of specialized training to master.[60] But several trends, such as the declining cost of DNA synthesis, the development of new techniques for DNA manipulation like CRISPR-Cas9, and the rise of computer-aided design tools and automation, all have the potential to lower these barriers.[61] Though the probability of such an event remains low, its potential impact could be very high. Combined with other factors, such as the continued threat of extremism in the region and the complexity of a response that must necessarily bridge national security and healthcare agencies, it is prudent for both countries to develop a playbook for this eventuality.[62]

---

[50] Singapore Ministry of Foreign Affairs. "Singapore - United States Third Country Training Program," August 4, 2018, http://www.mfa.gov.sg/Newsroom/Announcements-and-Highlights/2018/08/TCTPsigning.
[51] Cyber Security Agency of Singapore. "Singapore and the United States Sign Declaration of Intent on Cybersecurity Technical Assistance Program," November 16, 2018, https://www.csa.gov.sg/news/press-releases/singapore-and-the-us-sign-doi-on-cybersecurity-technical-assistance-programme.
[52] Gronvall, Gigi K. et al., "Singapore-U.S. Strategic Dialogue on Biosecurity" (Fort Belvoir, VA: Defense Technical Information Center, July 1, 2014), https://doi.org/10.21236/ADA612377.
[53] Inglesby, Tom, et al., "Southeast Asia Strategic Multilateral Biosecurity Dialogue: Meeting Report from the 2019 Dialogue Session" (Johns Hopkins Center for Health Security, May 2019), https://www.centerforhealthsecurity.org/our-work/publications/southeast-asia-strategic-multilateral-biosecurity-dialogue.
[54] Gronvall, et al., "Singapore-U.S. Strategic Dialogue on Biosecurity."
[55] One particularly compelling example of this is Japan, which managed to avoid massive outbreaks despite not instituting a lockdown because of Japanese scientists' understanding that Covid-19—like SARS—would feature indoor super-spreading events. See: Zeynep Tufekci, "This Overlooked Variable Is the Key to the Pandemic," *The Atlantic*, September 30, 2020, https://www.theatlantic.com/health/archive/2020/09/k-overlooked-variable-driving-pandemic/616548/.
[56] McArthur, Donna Behler. "Emerging Infectious Diseases," *The Nursing Clinics of North America* 54, no. 2 (June 2019): 297–311, https://doi.org/10.1016/j.cnur.2019.02.006. Brian McCloskey et al., "Emerging Infectious Diseases and Pandemic Potential: Status Quo and Reducing Risk of Global Spread," *The Lancet Infectious Diseases* 14, no. 10 (October 1, 2014): 1001–10, https://doi.org/10.1016/S1473-3099(14)70846-1.
[57] Coker, Richard J., et al., "Emerging Infectious Diseases in Southeast Asia: Regional Challenges to Control," *The Lancet* 377, no. 9765 (February 2011): 599–609, https://doi.org/10.1016/S0140-6736(10)62004-1.
[58] Fabian Koh, "Budget Debate: $90m to Be Spent on Singapore's First Top-Level Biosafety Lab, to Be Operational by 2025," *The Straits Times*, March 1, 2021, https://www.straitstimes.com/singapore/politics/90-million-to-be-spent-on-singapores-first-top-level-biosafety-lab-to-be.
[59] Gronvall et al., "Singapore-U.S. Strategic Dialogue on Biosecurity."
[60] Jefferson, Catherine, Filippa Lentzos, and Claire Marris, "Synthetic Biology and Biosecurity: Challenging the 'Myths,'" *Frontiers in Public Health* 2 (August 21, 2014), https://doi.org/10.3389/fpubh.2014.00115.
[61] Committee on Strategies for Identifying and Addressing Potential Biodefense Vulnerabilities Posed by Synthetic Biology. *Biodefense in the Age of Synthetic Biology* (The National Academies of Sciences, Engineering, and Medicine, 2018), https://doi.org/10.17226/24890.

*Biosecurity: existing U.S.-Singapore cooperation*

Compared with defense technology and cybersecurity, U.S.-Singapore cooperation on biosecurity is not as formally codified. While cybersecurity cooperation is addressed by, for example, the 2016 MOU and the 2018 agreement on regional technical assistance, there are no similar agreements on biosecurity. However, the United States and Singapore do participate in various multilateral forums such as the 70-country Global Health Security Agenda, launched in 2014 to bolster international capacity in combating infectious diseases.[63]

Singapore is also home to the U.S. Navy Medical Research Center—Asia (NMRCA), established in 2013 following the closure of its predecessor, Naval Area Medical Research Unit 2 (NAMRU-2) in Jakarta, Indonesia.[64] NMRCA manages a variety of projects across Southeast Asia, with a particular focus on research and surveillance of EIDs, and totaled 120 staff, including a detachment in Cambodia, as of 2019.[65] Like other cases, such as the 1990 MOU on the use of Singapore's naval facilities, NMRCA's relocation was the result of Singapore being willing to accommodate a U.S. military presence when other countries refused to do so. Political tailwinds in Indonesia led to the closure of NAMRU-2 in 2010, with the presence of U.S. Navy personnel deemed a non-starter, and NAMRU-2 staff were forced to relocate to Pearl Harbor until talks with Singapore were finalized.[66]

The United States and Singapore have also helped kick-start a Track II biosecurity dialogue in the region, led by the Johns Hopkins Center for Health Security (CHS). In 2014, CHS launched the dialogue as a U.S.-Singapore bilateral dialogue, but it has progressively expanded to include Indonesia, Malaysia, the Philippines, and Thailand as of 2019.[67] Though the dialogue is very much not a formal state-to-state Track I diplomatic effort, it nonetheless remains a valuable platform for regional biosecurity coordination given the dearth of similar dialogues and underscores how U.S.-Singapore biosecurity cooperation must necessarily take a holistic regional perspective over a purely bilateral approach.

**Future opportunities for collaboration**

Though the Enhanced DCA embraces cooperation in all three spheres, the general organizing principles for U.S.-Singapore cooperation in defense technology, cybersecurity, and biosecurity are all profoundly different. By its nature, defense technology cooperation is a particularly bilateral affair; it is also where the asymmetry of the relationship is most keenly felt. Despite the strides that Singapore's domestic defense industry has made, the chance of successful sales to the United States remains limited, and what sales there might be are far exceeded by Singapore's formidable appetite for cutting-edge U.S. platforms that it cannot produce itself.[68] Still, the two countries' shared interest in automation and robotics could, if pursued, lead to useful niche applications.

By contrast, U.S.-Singapore cooperation in cybersecurity and biosecurity—particularly the latter—is much better disposed to multilateral efforts. That makes their collaboration more regionally palatable and adds value to Singapore's contribution in its familiarity with Southeast Asian culture and politics and its ability to act as a hub for building regional capacity and networks. In cybersecurity, nation-state espionage campaigns will likely be a shared concern for the two countries, and they should improve both their own and other countries' defenses by prioritizing industrial sectors that nation-states frequently target, protecting against vectors of attack that nation-states frequently use. In biosecurity, the possibility of future pandemics arising elsewhere in Southeast Asia means that the United States and Singapore should put regional cooperation at the forefront and bolster epidemiological surveillance and response in other countries.

> **"In biosecurity, the possibility of future pandemics arising elsewhere in Southeast Asia means that the United States and Singapore should put regional cooperation at the forefront and bolster epidemiological surveillance and response in other countries. "**

*Defense cooperation: research and development*

Generally, among the most promising areas for bilateral collaboration in the defense technology space is the joint development of improved capabilities in C4ISR (i.e., command, control, communications, computers, intelligence, surveillance, and reconnaissance), particularly unmanned systems or systems with potential peacetime use.[69] On the U.S. side, there is a clear need for ISR (intelligence, surveillance, and reconnaissance) in particular: over two decades in the Middle East, the United States has enjoyed a largely unchallenged advantage in ISR,

---

[62] Inglesby et al., "Southeast Asia Strategic Multilateral Biosecurity Dialogue: Meeting Report from the 2019 Dialogue Session."
[63] Jenkins, Bonnie. "Now Is the Time to Revisit the Global Health Security Agenda," *Brookings Institution* (blog), March 27, 2020, https://www.brookings.edu/blog/order-from-chaos/2020/03/27/now-is-the-time-to-revisit-the-global-health-security-agenda/.
[64] U.S. Embassy in Singapore. "Naval Medical Research Center-Asia (NMRC-A)," accessed April 18, 2021, http://sg.usembassy.gov/embassy/singapore/sections-offices/naval-medical-research-center-asia-nmrc/. Sophal Ear, "Emerging Infectious Disease Surveillance in Southeast Asia: Cambodia, Indonesia, and the U.S. Naval Area Medical Research Unit 2," *Asian Security* 8, no. 2 (May 1, 2012): 164–87, https://doi.org/10.1080/14799855.2012.686338.
[65] U.S. Embassy in Singapore. "Naval Medical Research Center-Asia (NMRC-A)." U.S. Embassy in Singapore, "Fact Sheet: U.S.-Singapore Defense Cooperation."
[66] Ear, "Emerging Infectious Disease Surveillance in Southeast Asia." Doris Ryan, "Naval Medical Research Center – Asia Officially Opens Its Doors,"

*Naval Medical Research and Development News*, October 2013, https://upload.wikimedia.org/wikipedia/commons/d/d8/Naval_Medical_Research_and_Development_News_Vol_V_Issue_10_%28IA_NMRDNewsVolVIssue10%29.pdf.
[67] For more, see the CHS website, which provides meeting notes and additional materials for the dialogue, which has met about annually from 2014-19. In 2019, the participants additionally issued a joint statement: Anita Cicero et al., "Southeast Asia Strategic Multilateral Dialogue on Biosecurity," *Emerging Infectious Diseases* 25, no. 5 (May 2019), https://doi.org/10.3201/eid2505.181659.
[68] For example, Singapore proved unsuccessful in its attempt to sell Bionix vehicles to the United States in what would have been a $7 billion deal. See: Ron Matthews and Collin Koh, "Singapore's Defence-Industrial Ecosystem," in *The Economics of the Global Defence Industry*, ed. Keith Hartley and Jean Belin, 1st ed. (New York: Routledge, 2019), https://doi.org/10.4324/9780429466793.
[69] Cooper and Chase, *Regional Responses: Singapore*.

but it now faces the prospect of a major power conflict that could strip this advantage away.[70] Under such a scenario, kinetic attacks and electronic warfare could imperil U.S. air- and space-based ISR assets, which makes it important for the United States to retool its ISR platforms for greater robustness and survivability.

Robust C4ISR capabilities are similarly crucial for Singapore given its lack of strategic depth and reliance on technology as a force multiplier. There are some added considerations that encourage Singapore's investment in C4ISR rather than offensive capabilities: for one, acquiring certain capabilities could upset the regional balance of power, and hence prove counterproductive. For another, its small size relative to the United States means that it cannot collaborate meaningfully on larger, more expensive platforms. Combined, these mean that its contribution will likely skew toward the niche and defensive, but there remains plenty of room, particularly in C4ISR for such collaboration.

On defense technology, the two countries can consider investing in:

a. *Unmanned Systems.* Singapore's demographic decline—by 2030, it expects its annual number of conscripts to plummet to two-thirds from the late 2010s—means that unmanned systems are integral to its future defense planning, as a way to supplant its manpower shortfall.[71] It already has about 100 unarmed UAVs used for ISR but is considering renewing this aging fleet.[72] As it does so, it could consider working with the United States to investigate ways to harden unmanned systems against kinetic attacks and electronic warfare and possibly invest more in USVs and UUVs.

b. *Swarm Technology for ISR.* Rather than hardening expensive unmanned systems, one alternative that the United States and Singapore could pursue is swarm technology: that is, the deployment of multiple, individually inexpensive drones that work in tandem. As swarm technology has applications in disaster relief, the United States and Singapore could readily investigate it as part of their existing collaboration on AI for HA/DR. The lower cost of such drones may also make an investment in swarm technology better suited to Singapore's domestic defense industry.

c. *Anti-Drone Measures.* On the flip side, low-cost drones could also be used to gather intelligence on the United States or Singapore, as is already suspected to be happening with a series of alleged 'UFO' sightings near U.S. bases.[73] Swarms of such drones could even be used offensively against high-value assets in ways that would be costly or difficult to neutralize.[74] The two countries should hence collaborate on anti-drone technologies for both peacetime and wartime use. While kinetic weapons may be feasible in wartime, the risk of collateral damage means that in peacetime, the use of other tools, such as electromagnetic jamming, is preferable.

d. *Maritime Surveillance and Reconnaissance.* Though the rise of land-based hybrid warfare is well established, another prospect that the United States may want to guard against is the prospect of 'maritime hybrid warfare'—the disruption of maritime activities with deniable forces.[75] Singapore, being dependent on sea-lines of communication, shares these concerns if in a more general sense.[76] The difficulty of deterring such attacks makes it important to rapidly identify them with both onshore and offshore systems, so that they can be responded to in a timely fashion.[77]

e. *Data Fusion for Command and Control.* Data generated by ISR assets and other sensor instrumentation can be overwhelmingly heterogeneous, and requires processing to be useful for decision-making. Both the United States and Singapore are developing tools to integrate and interpret this data. The stated intent of the United States is to use AI to fuse disparate data sources into a 'common operating picture' for commanders.[78] Where possible, the two countries can pursue further collaboration on this front.

*Cybersecurity*

In cybersecurity, the United States and Singapore should bolster their defenses against advanced state actors, while also strengthening their societies and ASEAN member-states in general against financially motivated cybercriminals. To insulate themselves from nation-state actors, they need not explicitly identify their major concerns: rather, they can take measures to improve security in sectors that nation-states will likely target, warding against attack vectors that nation-states will likely use. At the same time, both countries should build on the work laid out in the 2016 MOU and their prior agreements on ASEAN-wide capacity building to increase the resilience of their respective countries and Southeast Asia as a whole.

Possible areas for cooperation include:

---

[70] Green, Michael, et al., *Asia-Pacific Rebalance 2025: Capabilities, Presence, and Partnerships* (Center for Strategic and International Studies, 2016).

[71] Koh, Eng Beng, "Preparing a Stout Defence for Generations to Come," *Pioneer Magazine*, August 1, 2019, https://www.mindef.gov.sg/web/portal/pioneer/article/feature-article-detail/ops-and-training/2019-Q3/aug19_fs1.

[72] Currently, Singapore's two largest platforms are the Israeli-made Hermes 450 and Heron 1 UAVs. Barry Desker and Richard A. Bitzinger, "Proliferated Drones: A Perspective on Singapore" (Center for a New American Security, June 2016), http://drones.cnas.org/reports/a-perspective-on-singapore/. Min Zhang Lim, "RSAF Tracking Developments in Drone Technology," *The Straits Times*, February 15, 2020, https://www.straitstimes.com/singapore/rsaf-tracking-developments-in-drone-technology.

[73] Rogoway, Tyler. "Adversary Drones Are Spying On The U.S. And The Pentagon Acts Like They're UFOs," *The Drive*, April 15, 2021, https://www.thedrive.com/the-war-zone/40054/adversary-drones-are-spying-on-the-u-s-and-the-pentagon-acts-like-theyre-ufos.

[74] Kuzma, Richard. "The Navy Littorally Has a Drone Problem," *War on the Rocks*, October 25, 2016, https://warontherocks.com/2016/10/the-navy-littorally-has-a-drone-problem/.

[75] Stavridis, James. "Maritime Hybrid Warfare Is Coming," *Proceedings, U.S. Naval Institute*, December 1, 2016, https://www.usni.org/magazines/proceedings/2016/december/maritime-hybrid-warfare-coming. Chris Kremidas-Courtney, "Countering Hybrid Threats in the Maritime Environment," *The Maritime Executive*, June 11, 2018, https://www.maritime-executive.com/editorials/countering-hybrid-threats-in-the-maritime-environment.

[76] Ang, Bertram Chun Hou. "Hybrid Warfare - A Low-Cost, High-Returns Threat to Singapore as a Maritime Nation," *Pointer, Journal of the Singapore Armed Forces* 44, no. 4 (2018): 26–37.

[77] Hawken, Colum. "Q-Boats and Chaos: Hybrid War on the High Seas," *RealClearDefense*, December 7, 2017, https://www.realcleardefense.com/articles/2017/12/07/q-boats_and_chaos_hybrid_war_on_the_high_seas_112748-full.html.

[78] Sayler, Kelley M. "Artificial Intelligence and National Security" (U.S. Congressional Research Service, November 10, 2020), https://fas.org/sgp/crs/natsec/R45178.pdf.

   a. *Supply Chain Security.* As demonstrated by Sunburst, software supply chain attacks—e.g., insertion of malicious code into third-party software updates—are particularly attractive to nation-state actors, as the large 'blast radius' of such attacks enables a well-resourced actor to compromise a large number of targets in a cost-effective way.[79] Though it can be more difficult to compromise a 'linchpin' third-party software vendor, the trust placed in such vendors makes such attacks difficult to guard against for downstream users. The United States and Singapore can jointly develop frameworks to manage software supply chain risks, taking inspiration from Singapore's scheme (announced early 2021) to incentivize critical infrastructure providers to do so.[80]
   b. *Maritime Cybersecurity Exercises.* In select industries where cybersecurity incidents could have transnational impacts and be geopolitically motivated, the two countries should conduct joint exercises and develop a playbook for a bilateral response, or even a multilateral one if they included other ASEAN countries.[81] One prime candidate is the maritime sector, which could become a target for nation-state actors due to geopolitical reasons, but lags other sectors, such as the financial sector, in terms of cybersecurity practices and resources.[82]
   c. *Operational Technology Security.* Operational technology (OT) and information technology (IT) are typically contrasted because the former is used in industrial operations and the latter in an administrative context. OT's industrial role means that OT cyber incidents can have a physical impact, such as manufacturing disruptions or even loss of life.[83] From the geopolitical perspective, OT systems can therefore be attractive targets. In recent years, the increasing number of OT attacks has been further exacerbated by a growing tendency to connect OT systems to Internet-facing systems and networks.[84] The potential high impact of such incidents means that Singapore and the United States should share best practices on classifying and managing OT systems, particularly as the OT vs. IT line continues to blur.
   d. *Regional Capacity Building.* The United States and Singapore should continue building ASEAN's cybersecurity capacity through the TCTP and related programs, and expand these if possible. In addition, the United States can support the ASEAN-Singapore Cyber Center of Excellence by lending expertise, or possibly even with financial contributions. Where possible, both countries should engage other ASEAN countries in their other bilateral efforts, such as on software supply chain security.

*Biosecurity*

As one Singaporean participant mentioned in the 2014 U.S.-Singapore CHS dialogue, 'a chain is only as strong as its weakest link,' and biosecurity in the two countries depends heavily on other weak links they may be connected to.[85] Capacity building to boost regional pandemic preparedness in Southeast Asia should hence be the key focus of U.S.-Singapore cooperation on biosecurity. To enable the timely detection of new EIDs, the two countries should invest in both technical and human resources to build up regional epidemiological surveillance networks. At the same time, as both countries position themselves to participate in the synthetic biology economy, they should prioritize biosafety regulations.

Possible areas for cooperation include:
   a. *Technology for Rapid Testing and Reporting.* Rapid diagnostic test kits, which provide on-the-spot results without samples being sent to a lab, are useful for epidemiological surveillance in lower-income countries.[86] However, many such tests are lower in accuracy and are no substitute for lab diagnostics.[87] Singapore and the United States could fund the development of high-accuracy point-of-care diagnostic tests for use in Southeast Asia, as private sector companies have limited incentive to develop such tests otherwise. They should also invest in tools that can integrate this decentralized test data for better decision-making during rapidly evolving outbreaks.[88]
   b. *Knowledge Exchange for Epidemiological Surveillance.* However, technology alone does not guarantee effective epidemiological surveillance in the whole Southeast Asia. Political and economic factors can hamper the work of local professionals or deter them entirely, creating a manpower shortage.[89] Many such factors are domestic, and hence beyond the power of external parties to address. Still, the United States and Singapore should continue existing training programs like the TCTP and build Singapore into a hub for regional knowledge exchange. In parallel with the ASEAN-Singapore Cybersecurity Center of Excellence, Singapore could consider establishing a similar regional 'center of excellence' for biosecurity, bringing together its maritime neighbors to complement the existing Mekong Basin Disease Surveillance Network.

---

[79] Herr, Trey, et al., "Breaking Trust: Shades of Crisis across an Insecure Software Supply Chain" (Atlantic Council, July 26, 2020), https://www.atlanticcouncil.org/in-depth-research-reports/report/breaking-trust-shades-of-crisis-across-an-insecure-software-supply-chain/. While a full discussion of software supply chain security is beyond the scope of this paper, the Herr et al. (2020) report provides several excellent recommendations on this topic.
[80] Chee, Kenny. "Push to Better Manage Cyber-Security Risks in Critical Infrastructure," *The Straits Times*, March 3, 2021, https://www.straitstimes.com/singapore/push-to-better-manage-cyber-security-risks-in-critical-infrastructure.
[81] Manantan, Mark and Eugenio Benincasa, "U.S.-Singapore Cyber & Tech Security Virtual Series Session #4: U.S.-Singapore Perspectives on Enhancing Critical National Infrastructure Cybersecurity," *Pacific Forum* (blog), February 4, 2021, https://pacforum.org/events/us-sg-cybertech-4.
[82] Burlend, William and Jack David, "'Resilient Seas' - Cyber Security Threats to the Maritime Industry," *PricewaterhouseCoopers* (blog), January 20, 2020, https://pwc.blogs.com/cyber_security_updates/2020/01/resilient-seas-cyber-security-threats-to-the-maritime-industry.html.

[83] Lakhani, Aamir. "Evolution of Cyber Threats in OT Environments," *Fortinet* (blog), June 11, 2020, https://www.fortinet.com/blog/industry-trends/evolution-of-cyber-threats-in-ot-environments.html.
[84] "Singapore's Operational Technology Cybersecurity Masterplan 2019" (Cyber Security Agency of Singapore, October 2019), https://www.csa.gov.sg/news/publications/ot-cybersecurity-masterplan.
[85] Gronvall, et al., "Singapore-U.S. Strategic Dialogue on Biosecurity."
[86] Kozel, Thomas R. and Amanda R. Burnham-Marusich, "Point-of-Care Testing for Infectious Diseases: Past, Present, and Future," *Journal of Clinical Microbiology* 55, no. 8 (August 1, 2017): 2313–20, https://doi.org/10.1128/JCM.00476-17.
[87] Tan, Audrey. "Rapid Covid-19 Test Kits Not Used in Singapore as They Can Miss True Cases," *The Straits Times*, September 18, 2020, https://www.straitstimes.com/singapore/health/rapid-test-kits-not-used-here-as-they-can-miss-true-cases.
[88] Ming, Damien, et al., "Connectivity of Rapid-Testing Diagnostics and Surveillance of Infectious Diseases," *Bulletin of the World Health Organization* 97, no. 3 (March 1, 2019): 242–44, https://doi.org/10.2471/BLT.18.219691.
[89] Ear, "Emerging Infectious Disease Surveillance in Southeast Asia."

c. *Journalism Fellowship Program.* The spread of misinformation during Covid-19 shows that effective risk communication is vital to pandemic management. To improve this, Singaporean and U.S. stakeholders could jointly launch a health journalism fellowship in Southeast Asia, to help participants build a network of scientific and professional resources that they can tap into their work. There are existing analogs for this: in Singapore, the Temasek Foundation and Institute of Policy Studies run an 'Asia Journalism Fellowship,' while the East-West Center, Stimson Center, and Internews Earth Journalism Network jointly run a 'Mekong Data-Journalism Fellowship.'[90]

d. *Biosafety Regulations in Synthetic Biology.* Singapore has a reputation for regulatory innovation, being willing to work closely with industry and being agile due to its size. As it invests in synthetic biology, it could coordinate with U.S. agencies to explore current proposals for improving biosafety in synthetic biology, such as mandatory screening for third-party DNA synthesis, which has not been implemented at the federal level in the United States. This could facilitate fine-tuning of innovative regulatory proposals, balancing biosafety and innovation, that could in the future be implemented at the state or federal level in the United States.

---

[90] Temasek Foundation and Institute of Policy Studies. "Overview," Asia Journalism Fellowship, May 21, 2015, https://www.ajf.sg/overview/. Katie Bartels, "Mekong Data-Journalism Fellowship," East-West Center, March 21, 2019, https://www.eastwestcenter.org/professional-development/mekong-journalism-reporting-fellowship.